# Machine Learning in Ethnobotany

Marc Böhlen
*Department of Art
Emerging Practices in
Computational Media
University at Buffalo*
Buffalo, USA

marcbohlen@protonmail.com

Wawan Sujarwo
*Ethnobiology Research Group
Research Center for Biology
Indonesian Institute of Sciences*
Cibinong, West Java, Indonesia

wawan.sujarwo@lipi.go.id

*Abstract*— We describe new opportunities created by bring A.I. to the field of ethnobotany. In particular we describe a novel approach to ethnobotany documentation that harnesses machine learning opportunities, specifically for the documentation of traditional ecological knowledge with mobile phones in emerging economies. Using a case study on the island of Bali as a departure point, the project maps out machine learning approaches to documentation and responds to technology and capital gradients between research contexts in the global north and south in an attempt to capture knowledge that might otherwise not be represented.

*Keywords—ethnobotany, machine learning, knowledge representation, convolutional neural networks, local ecological knowledge, emerging economies*

## I. INTRODUCTION

Ethnobotany is the scientific study of traditional knowledge of people and plants, ranging from medical to religious use. In general, ethnobotanical data are collected *manually* in the field through discussions with key informants and semi-structured interviews [2]. Informants are typically asked to provide a list of plants (with any use-categories) grown and consumed in their village.

This project's goal is to investigate the potential of Machine Learning (ML) to support existing ethnobotanical research methods and to explore new methods ML can offer to ethnobotany. More specifically we want to probe whether ubiquitous mobile media can serve as a viable data collection vector, identify which types of ML operations can process such data and determine what kinds of insights this combination might generate. In order to get a practical understanding of the challenge, we create a test site in Central Bali, Indonesia with a rich history of ethnobotanical practices.

## II. ETHNOBOTANY IN BALI

Despite a surface area of only 5,577km$^2$, Bali possesses a rich biological diversity, including some 1,595 species of Spermatophyta, 173 species of Pteridophyta, and 169 species of Bryophyta [12]. Moreover, Bali is an extremely interesting area for ethnobotany as local knowledge is preserved in Balinese daily life, including in rituals, traditional medicines, and local foods. Over the last decade, many ethnobotanical studies in Bali have focused on the use of medicinal and food plants [18], [20]. One ethnobotanical study carried out an evaluation of ritual plants [19] and a few ethnobotanical studies have carried out quantitative evaluations of biocultural diversity on the island [1], [18]. At the same time, cultural erosion, particularly of the knowledge of Bali's edible plants, appears to be on the rise [21]. This decline in traditional ecological knowledge is a prime motivator for our project.

## III. MACHINE LEARNING IN ETHNOBOTANY

### A. Inspiration

In the 1930s, social anthropologists Margaret Mead and Gregory Bateson combined photography, film and field notes in an attempt to capture the culture of Central Balinese peoples in Bayung Gede [4]. Mead and Bateson sought a method to comprehensively represent the culture under inquiry. The researchers understood that synergies across distinct documentation techniques could capture fine grained information that any single method performed in isolation could not. And Mead in particular recognized that filmmakers and artists were differently enabled to contribute to this opportunity [7].

### B. Machine learning for species identification

There is a substantial amount of recent research on the automated identification of species. From a technical standpoint, the research on Convolutional Neural Network (CNN) architectures for species identification is relevant (fish [3]; insects [14], [23]; plankton [11]; and plants [9], [10], [13], [15], [16], [17], [22]). Of particular interest are: [23], an approach to taxonomic identification of small datasets of insects with a network that was pre-trained on generic data with feature transfer; [11], the use of contextual metadata to support convolutional neural nets in their ability to classify plankton; [16], experiments with various field collection methods and image pre-processing strategies used to improve classification. Furthermore [17], an approach to image-based hierarchical plant classification with images in the wild; [22], the 2018 Expert LifeCLEF approach to automated plant species identification with fine-tuned high-dimensional CNNs; and [13], an account of the 2019 LifeCLEF plant identification challenge that sought to address classification of data deficient tropical plants.

*C. Beyond species identification*

None of the technical approaches described above combine ethnographic or plant use data together with species classification; all of these systems are designed for scientists and botanists with little concern for the constituents who live with the life forms encountered. Inspired by Mead and Bateson, we want to determine if and how ML might contribute to a deeper representation of lived botany and the representation of local ecological knowledge.

## IV. FIELD STUDY

*A. Labelling data with mobile phones*

We approach the practicality of ML in ethnobotany from the perspective of knowledge representation. Because mobile phones are ubiquitous across emerging economies, we use mobile phones as data capture devices. Moreover, mobile phones allow us to respond to a central challenge of supervised ML, namely the need for large and high quality datasets as previous research in species identification has shown [4], [8].

The use of mobile phones for data collection is facilitated in our approach through a recent invention[1]. The approach makes use of the synchronicity between streaming imagery and audio tracks in video formats to create labeled images from video collected in the field, allowing large numbers of labeled images to be created from limited field expeditions. Unlike other methods [16], this one requires no manual post processing such as cropping and segmentation, substantially reducing the effort required to produce high quality image data. Furthermore our approach allows videos to be post-processed in the lab with annotations by experts not present during data collection in the field. This approach allows anyone with a mobile phone to contribute to data creation, and also reduces the barriers to data set expansion as recognized data deficiencies can be more efficiently targeted with additional information collected at a later time.

*B. Field study sites in Central Bali*

We have established several nodes in Central Bali where collaborators collect data on plant occurrences and usage. One node is in Kerta village, a second one in Bukian village and a third one in Sekaan village. At each site one or more local experts are responsible for documenting plants of ethnobotanical significance in consultation with our research team. In Kerta village, for example, Made Darmaja maintains a forest-garden with dozens of tropical plants and fruits, several of which are indigenous to Bali. At Sekaan village, Gusti Sutarjana and his family have maintained a forest garden under family care for over seven generations. At the time of this writing, the data collection teams have collected hundreds of field videos from which over 50'000 images of 26 of some of Bali's most ethnobotanically significant plants have been generated through the approach outlined above.

## V. BALI-26 DATASET

The Bali-26 data set[2] includes Aroid *(Amorphophallus paeoniifolius)*, Bamboo Petung (*Dendrocalamus asper*), Cacao (*Theobroma cacao*), Coffea Arabica (*Coffea canephora*), Indonesian Cinnamon (*Cinnamomum burmanii*), Durian (*Durio zibethinus*), Dragonfruit (*Hylocereus costaricensis*), Frangipani (*Plumeria alba*), Guava (*Psidium guajava*), Jackfruit (*Artocapus heterophyllus*), Lychee (*Litchi chinensis*), Mango (*Mangifera indica*), Mangosteen (*Garcinia mangostana*), Nilam (*Pogostemon cablin*), Papaya (*Carica papaya*), Passiflora (*Passiflora edulis*), Sawo (*Manilkara zapota*), Snakefruit (*Salacca zalacca*), Starfruit (*Averrhoa carambola*), Sugar Palm (*Arenga pinnata*), Taro (*Colocasia esculenta*), Vanilla (*Vanilla planifolia*), Water Guava (*Syzygium aqueum*), White Pepper (*Piper nigrum*) and Zodia (*Evodia sauveolens*).

The data set includes examples from multiple plants of each category and from multiple facets of each of the plants, including leaves, flowers, fruits and bark, where applicable. While this broad documentation approach produces a rich visual representation of the target flora, it also uncovers how much more information one could add to the plant collection. Moreover, several of our plant categories have multi-faceted appearances and yet are grouped under a single category for ease of classification. A case in point is the dragon fruit. Figure 1 shows seven different morphologies of the dragon fruit: as a ripe leaf, a young leaf, a flower, and a fruit in various stages of ripeness. As dragon fruit is a climber, it uses many different structural support options – typically other plants - and thus always appears in a different context.

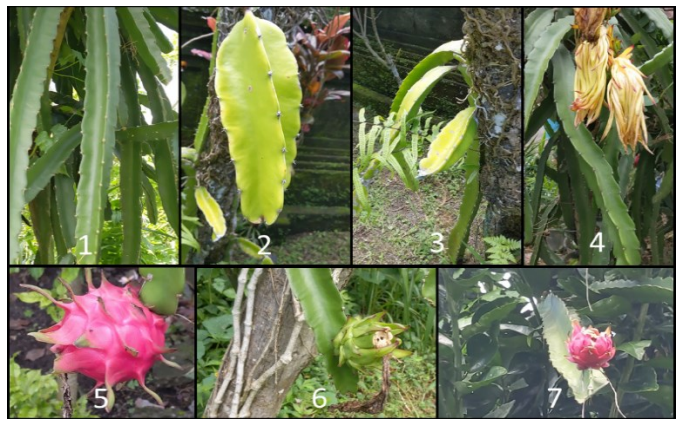

Fig.1. Examples of dragon fruit in the Bali-26 dataset; 1 and 2 leafs; 3 and 4 flowers; 5, 6 and 7 fruit in various stages of ripening.

---

[1] University at Buffalo, Office of Research and Economic Development. New Technology Disclosure 030-7278: Expertise collection with action cameras. Filed by Marc Böhlen, October 23, 2018.

[2] https://filedn.com/lqzjnYhpY3yQ7BdfTulG1yY/bali-26.zip

Moreover, the collection is somewhat uneven; not all fruits were in season during our first data collection effort. There can be no doubt that the 50,000+ images from the Bali-26 data set cover only a small segment of ethnobotanically relevant species and conditions; it has no claim of completeness and serves us here only as a proof of concept dataset. Nonetheless, Bali-26 offers ML a multiplicity of challenges as the following sections illustrate.

## VI. Enabling software: Catch and Release

In order to facilitate research activities across our team members, we have developed an intuitive interface to the code base required to process video feeds to labeled images: *Catch & Release* (C&R)[3]. C&R is designed to allow non-computer specialists from various disciplines to participate, and to experiment with data collection and preparation for supervised machine learning. It allows researchers to pre-process video streams, analyze interviews and create customized image sets as inputs for a variety of state-of-the-art neural network based classifiers.

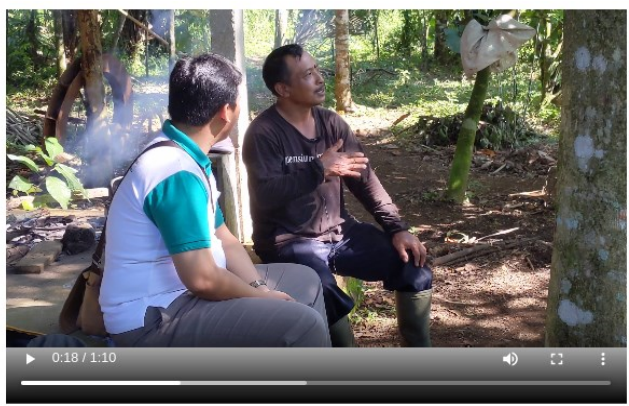

Fig. 2. Interview processing module of C&R.

### A. Interview to text

C&R allows ethnobotanists to process video interviews; the most common method of capturing nuanced information in local ecological knowledge of individuals. In our test case, we performed an interview with Made Darmaja, the owner of a forest garden in Kerta village (Fig. 2). Darmaja is a unique individual with a vast body of knowledge and practical insights into the interdependencies of various plants, insects and animals in his forest. We used a mobile phone to record the interview in Bahasa Indonesia and processed the resultant video with C&R to extract text from video. While the accuracy of the video to text system, currently based on a commercial Speech API, is far from perfect, it allows one to rapidly scan through a field recording. Because the system lets us specify confidence levels, we have a handle on the imperfection that even state of the art transcription systems contain. Moreover, the system has language processing abilities that allow for keyword search and the ability to relate results to specific moments in the interview. To be clear, there is no intention to replace or compromise the richness of original field recordings or long interviews with these automated transcriptions. However, this feature can serve as an accelerator and detect correlations in time-based data that might elude a human ethnographer.

### B. Image classification

The Bali-26 image set was produced with our video-to-labeled-image system and applied to five convolutional neural network architectures (both untrained and pre-trained): Alexnet, Squeezenet, Resnet50, Resnext152 as well as a low-dimensional generic CNN (Vanillanet) as a baseline. Prior to network training, the image set was shuffled, balanced and normalized. Each network was trained over 50 epochs with training and evaluation sets divided equally (50:50). An automated image quality routine removed excessively fuzzy images, and we manually removed out-of-context images from the beginning and end of some of the video sequences. Input images were then processed on single GPU computers with a PyTorch framework under Linux (Ubuntu).

## VII. Evaluation

Table 1 provides an overview of the responses of the five selected convolutional neural nets to the Bali-26 data set. The simpler Vanillanet, Alexnet and Squeezenet models suffer from substantial top 1 error rates while the high-dimensional Resnext50 and Resnet152 networks perform much better. Resnet152 even performs on Bali-26 just as well as it does on the canonical ImageNet[4] data set. In our case, we observe substantial differences across categories, however (see Fig. 3). Some categories such as zodia and nilam have very low error rates while others such as mango and water guava have much higher error rates. An additional factor likely affecting the error spread is the ease with which videos can be collected and converted to labeled images, potentially creating oversampled image collections. We have partially addressed this downside of the data collection approach by limiting the number of frames-per-second in the conversion from video to image and by and balancing the distribution of images across all categories such that each category contains at least 1200 and no more than 2500 examples.

---

[3] https://github.com/realtechsupport/c-plus-r.git

[4] http://www.image-net.org/

Furthermore, the convenience and informality of mobile media video collection challenges the controlled image capture that single image photography allows for, and the widely varying image conditions across the videos of leafs, tree bark, flowers and plants, etc. adds to the informational complexity of the dataset. In some cases, our data collection efforts coincided with low occurrences of certain plants. For example, in Bali the mango season runs from November to January and the lychee season occurs mostly in December. Our February-March data collection was not able to adequately represent the richness of these two important categories and the classifier performance reflects these deficiencies.

Additional experiments to improve data quality and network performance are ongoing, but not part of the main line of arguments in this text and are not discussed here for that reason.

| CNN model | top-1 error on ImageNet [5] | top-1 error on Bali-26; pretrained | best training accuracy on Bali-26; pretrained |
|---|---|---|---|
| Vanillanet | n.a | 57 | 92 |
| Alexnet | 43 | 47 | 97 |
| Squeezenet | 42 | 62 | 98 |
| Resnext50 | 22 | 28 | 99 |
| **Resnet152** | **22** | **21** | **99** |

Table 1. Comparative performance of CNN architectures on the Bali-26 data set. Top error values averaged (percentages) across all categories.

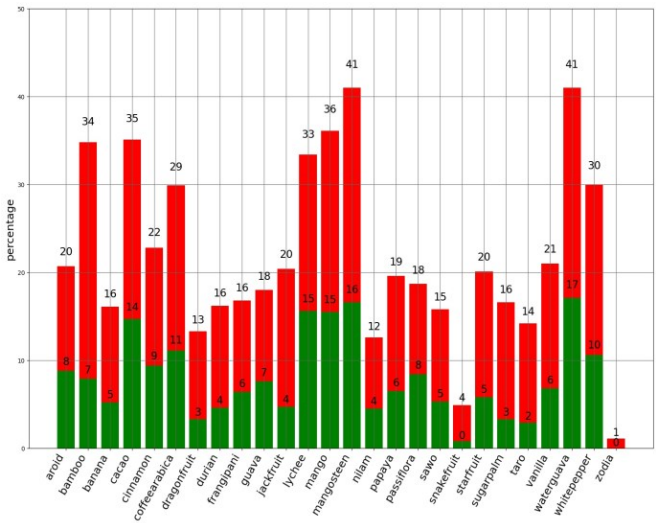

Fig. 3. Resnext152 performance: Top-1 and top-3 error rates across all categories of the Bali-26 data set.

## A. Image classification across contexts

Ethnobotany seeks to understand the varied uses of plants. Consequently, ethnobotanists document plant use in a variety of contexts and locations. Our ML supported approach has demonstrated unique capabilities that can assist in this documentation process.

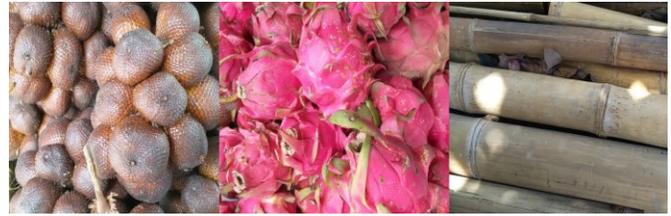

Fig.4. Left: Snake fruit at a market. Center: Dragon fruit at a market. Right: Bamboo at a construction site.

The three images in Figure 4 show snake fruit, dragon fruit and bamboo in daily use scenarios that constitute only about 10% of the respective image collections. Without additional information, the best performing classifier, Resnet152, can robustly identify these images in different contexts as listed in table 2.

| Resnet152 | snake fruit (market) | dragon fruit (market) | bamboo (construction) |
|---|---|---|---|
| **top-1 error** | **38** | **2** | **8** |
| top-3 error | 2 | 0 | 0 |

Table 2. Resnext50's ability to recognize 3 image categories in varied use contexts as described above. Tests performed on 70 to 85 images depicting these contexts from a test set (not training set) collected from the source videos. If the same network is trained on images with zero out-of-context examples, then the error rates increase markedly.

This recognition across contexts is due to the ability of a well-trained classifier to generalize and to identify similarities across widely varying conditions. It provides an interesting and to date unused asset for ethnobotany research while also demonstrating limitations as the classifiers cannot identify the contexts themselves.

## VIII. EXPANDING THE COLLECTION

Because of our investment in creating a small community of plant observers and data collectors, and because C&R can easily process additional video feeds, we are in the position to be able to add new data to the existing collection. We have tested this secondary data collection option on cinnamon, as we were not able to collect sufficient examples of cinnamon during the first expedition. In a second data collection effort, one of our local team members located cinnamon plants in various stages of growth, documented them via mobile phone, and sent the videos to the research team for processing. The result is a markedly improved performance of the classifiers. For example, the top-1 error of Resnext50 for Indonesian cinnamon improved from 26% to 5% with images from ten twelve second videos. However, the expansion of the cinnamon collection

---

[5] https://pytorch.org/docs/stable/torchvision/models.html

negatively impacted the lychee category, already far from ideal as mentioned above.

## IX. CONTRIBUTIONS TO THE FIELD

### A. Rethinking the concept of data in the wild

We understand now that collecting data sets from multiple instances of a plant category in different contexts, locations, lighting and environmental conditions is key to robust data set creation and a prerequisite for the ability of an image classifier to generalize. Independent of our system's ease of data collection, all data sets for machine learning must be carefully curated; C&R and its first product Bali-26 is subject to the same constraints. However, C&R has the advantage of allowing us to address data deficiencies and to improve the data set over time without excessive effort, allowing us to respond to the fact that the plants - and some of the uses they are put to - vary across seasons. As such, C&R contributes to a viable framework for richer ethnobotany data sets and subsequent investigations thereof.

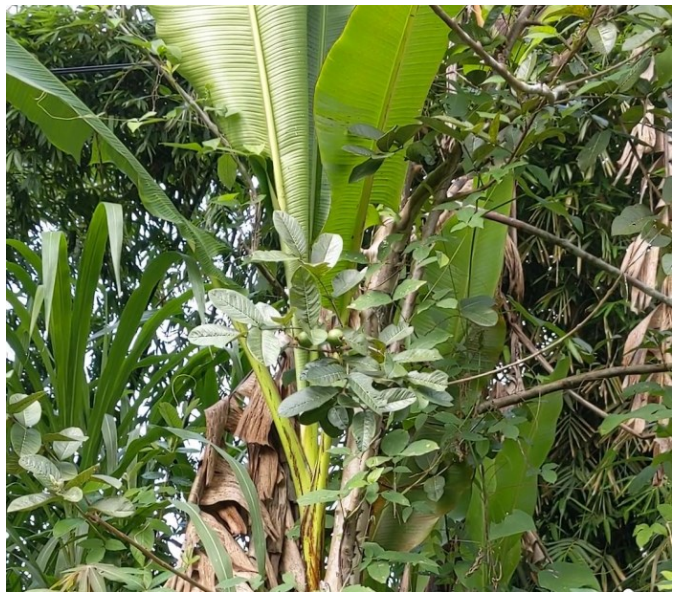

Fig. 5. View of a forest in Central Bali: A tangled mix of bamboo, banana, elephant grass, and guava.

In computer science, images 'in the wild' are images that are not laboratory generated. However, they are almost never wild and always engineered in one way or another - and more often than not simply scraped from the Internet. Even our most controlled images will be classified by computer science standards as images in the wild. While our videos were collected in the forests of Central Bali, the images we generate from them focus on individual plants in varying contexts. This focus on individual plants makes the task of the classifiers more tractable but ultimately reduces the complexity and richness of the plants and forests. Figure 5 gives an impression of how the plants appear in a tropical forest such as Central Bali; often tangled and mixed in place with instances of isolated plants the exception. The use of ML in ethnobotany can provide an opportunity to think more deeply about wild images, wild things and the many forms of reduction that occur when researchers produce input to classification systems. And it might yield an opportunity to devise approaches that can more robustly represent this level of richness.

### B. Considering qualitative dimensions of classification

We have yet to develop a method by which we can combine information collected from the interview module with the image classifiers. Related research [11] suggests that quantitative meta-data (such as plant size or weight) can be incorporated into classifier architectures to augment image based classifiers to cross-domain classifiers. However, it is not clear how *qualitative* information might be included in such an operation. How can we include the ritual importance of a given plant, or the significance of a tree in family possession for generations? As such, this project points to new opportunities for transdisciplinary research in A.I. enabled field study methods.

### C. Countering technology dependencies

Our approach to ML in ethnobotany addresses several dimensions of technology dependency. First, the C&R code base is open source and available on GitHub. Second, we invested efforts to ensure that the barriers to system use for constrained research budgets are low by repurposing used business laptops with an open source operating system. Such considerations are essential when working across technology and capital gradients that exist between established and emerging economies [5], [6].

While the current video-to-label system does connect to a commercial API, our approach is designed to function at no cost, as the audio data are sliced into small packets that fall under the billing threshold. We have created an open source alternative for speech processing[6] in interviews, but we cannot apply it to the Bali use case as we do not have access to a sufficiently developed Bahasa Indonesia language corpus. Furthermore, we trained all our classifiers de novo on the Bali-26 data set directly with lukewarm results (see Table 1). With the exception of the simplest and least powerful classifier (Vanillanet) all our neural nets performed better with prior training on ImageNet. This is a dependency that we wish to remove in the future as ImageNet is compromised by bias in its image set curation.

Our data contains location specific information which we choose to make no programmatic use of. While some traditional Bali villages already protect rare and useful flora [24], C&R might be used in other contexts with fewer controls in place in the future. By not including a mapping ability into C&R, we prevent the spread of all too easy location data that can be misused and harmful in ecologically delicate environments.

---

[6] https://cmusphinx.github.io/

## D. Describing underrepresented forms of knowing

This project allows us to represent knowledge that would otherwise not be captured in a form amenable to state of the art supervised learning systems. As such our project contributes to expanding the scope of knowledge that ML systems and all their downstream dependencies process and distribute; it allows at least some of Bali's rich ethnobotanical knowledge to be represented in the 21st century canonical and technical epistemology of machine learning. And because C&R is available to other researchers, new repositories from established and emerging economies capturing maybe even less accessible information might become accessible to ML systems.

## X. CONCLUSION AND FUTURE WORK

### A. Metadata, scenario recognition, voting

We are investigating better methods of preventing overfitting in the abundance of image data produced by our videos. Additionally we intend to explore the potential of including meta-data collected from non-image sources, such as interviews or remote sensing modules into our classification nets. We also want to explore hierarchical labeling within our existing categories - and related to this more common option - the ability to differentiate use scenarios (Banana as a daily food versus Banana as an offering during the *Panca Yadnya* rituals, for example). We intend to perform experiments with more sophisticated classifiers and will investigate the random-forest inspired option of voting to determine the best outcome of a classification task, seeking a majority consensus across multiple classifiers instead of blindly relying on a single and possibly compromised classifier.

### B. Beyond Bali-26

Our goal is to create a much larger representation of Bali's ethnobotanically significant resources. A substantially enlarged collection would likely allow us to train all networks de novo with improved performance. Moreover, we would like to expand the effort to other regions with underrepresented knowledge forms. ML databases suffer from a lack of diversity and richness. For this reason, it could be wise to have classifiers train on data more in tune with the lush complexity of the world. Richer ML and more active participation in the application of ML from peoples across the world would be a desirable outcome of such an effort.

ACKNOWLEDGMENT

Classifier training was supported in part by a Google faculty research grant.

REFERENCES

[1] A. Agung, "Bali endangered paradise? Tri Hita Karana and the conservation of the island's biocultural diversity", Universiteit Leiden Press, Leiden, 2005.

[2] M. N. Alexiades, and J. W Sheldon, "Selected guidelines for ethnobotanical research: A field manual", The New York Botanical Garden Press, New York, 1996.

[3] V. Allken, N. O. Handegard, S. Rosen, T. Schreyeck, T. Mahiout, and K. Malde, "Fish species identification using a convolutional neural network trained on synthetic data", ICES J. Mar. Sci. 76 2019, no. 1, pp. 342-349.

[4] G. Bateson and M. Mead, "Balinese character; a photographic analysis", New York Academy of Sciences, 1942.

[5] M. Böhlen, I. Maharika, Y. Ziyan, Y. and I. Hakim, "Biosensing in the Kampung", Intelligent Environments. IEEE Computer Society, Washington, DC, USA, 2014, pp. 23–30.

[6] M. Böhlen, and I. Maharika, "Cloud computing in the Kampung", Hybrid City: Data to the People, 2015, URIAC, Athens.

[7] S. Brand, "For God's Sake Margaret! Conversation with Gregory Bateson and Margaret Mead", CoEvolutionary Quarterly, 10 (21), pp. 32-44. June 1976.

[8] G. Caneva, L. Traversetti, W. Sujarwo, and V. Zuccarello, "Sharing ethnobotanical knowledge in traditional villages: Evidence of food and nutraceutical core groups in Bali, Indonesia", Economic Botany 71, 2017, no. 4, pp. 303-313.

[9] J. Carranza-Rojas, A. Joly, H. Goëau, E. Mata-Montero, and P. Bonnet, "Automated identification of herbarium specimens at different taxonomic levels", Multimedia Tools and Applications for Environmental & Biodiversity Informatics (Alexis Joly, Stefanos Vrochidis, Kostas Karatzas, Ari Karppinen, and Pierre Bonnet, eds.), Springer International Publishing, Cham, 2018, pp. 151-167.

[10] M. Dyrmann, H. Karstoft, and H. S. Midtiby, "Plant species classification using deep convolutional neural networks", Biosystems Eng. 151, 2016, pp. 72-80.

[11] J. S. Ellen, C. A. Graff, and M. D. Ohman, "Improving plankton image classification using context metadata", Limnol. Oceanogr. Methods, 17-8, 2019, pp. 439-461.

[12] D. Girmansyah, Y. Santika, A. Retnowati, W. Wardani, I. Haerida, E. Widjaja, M. van Balgooy, "Flora of Bali: An annotated checklist", Yayasan Pustaka Obor, Jakarta, 2013.

[13] H. Goeau, P. Bonnet, and A. Joly, "Overview of LifeCLEF plant identification task 2019: diving into data deficient tropical countries", 2019.

[14] O. Hansen, J. Svenning, K. Olsen, S. Dupont, B. H. Garner, A. Iosfidis, B. W. Price, and T. T. Hoye, "Species Level image classification with convolutional neural network enables insect identification from habitus images", Ecol. Evol. 39, 2019, pp. 737-747

[15] H. Hiary, H. Saadeh, M. Saadeh, and M. Yaqub, "Flower classification using deep convolutional neural networks", Institution of Engineering and Technology 12, no. 6, 2018, pp. 855-862.

[16] M. Rzanny, M. Seeland, J. Wäldchen, and P. Mäder, "Acquiring and preprocessing leaf images for automated plant identification: understanding the tradeoff between effort and information gain", Plant Methods 13, 97 (en), 2017, pp. 1-11.

[17] M. Seeland, M. Rzanny, D. Boho, J. Wäldchen, and P. Mäder, "Image based classification of plant genus and family for trained and untrained plant species", BMC Bioinformatics 20, no. 1, 4, 2019.

[18] W. Sujarwo, I. B. Arinasa, G. Caneva, and P. Guarrera, "Traditional knowledge of wild and semi-wild edible plants used in Bali (Indonesia) to maintain biological and cultural diversity", Plant Biosystems - An International Journal Dealing with all Aspects of Plant Biology 150, no. 5, 2016, pp. 971-976.

[19] W. Sujarwo, G. Caneva, and V. Zuccarello, "Patterns of plant use in religious offerings in Bali (Indonesia)", Acta Botanica Brasilica 34 no. 1, 2020, pp. 40-53.

[20] W. Sujarwo, and G. Caneva, "Using quantitative indices to evaluate the cultural importance of food and nutraceutical plants: Comparative data from the island of Bali (Indonesia)", Journal of Cultural Heritage, no. 18, 2016, pp. 342–348.

[21] W. Sujarwo, I. B. Arinasa, F. Salomone, G. Caneva, and S. Fattorini, "Cultural erosion of Balinese indigenous knowledge of food and nutraceutical plants", Economic Botany 68, no. 4, 2014, pp. 426–437.

[22] M. Sulc, L. Picek, and J. Matas, "Plant recognition by inception networks with test-time class prior estimation", CLEF, 2018.

[23] M. Valan, K. Makonyi, A. Maki, D. Vondracek, and F. Ronquist, "Automated taxonomic identification of insects with Expert-Level accuracy using effective feature transfer from convolutional networks", Syst. Biol. 68,, no. 6, 2019, pp. 876-895.

[24] N. Wijana, I. Setiawan, "Rare Plant Preservation through Village Forest Policy in Bali", Proceedings of the 2nd International Conference on Innovative Research across Disciplines. Atlantis Press, 2017.